\documentclass[aps,pra,twocolumn,groupedaddress]{revtex4-1}
\usepackage{amsmath}
\usepackage{graphicx}
\usepackage{amsfonts}
\usepackage{color}

\usepackage{dcolumn}
\usepackage{bm}
\usepackage{multirow}
\usepackage{array}
\usepackage{booktabs}
\usepackage{ctable}
\usepackage{upgreek}
\usepackage{epsfig}
\usepackage{mathrsfs}
\usepackage{amssymb}
\usepackage{amsbsy}
\usepackage{cancel}
\usepackage{pifont}
\usepackage{marginnote}
\usepackage{float}
\makeatletter

\begin{document}

\relax

\title{Non-equilibrium phase diagram of a 1D quasiperiodic system with a single-particle mobility edge}

\date{\today}

\author{Archak Purkayastha}
\affiliation{International centre for theoretical sciences, Tata Institute of Fundamental Research, Bangalore - 560089, India}
\author{Abhishek Dhar}
\affiliation{International centre for theoretical sciences, Tata Institute of Fundamental Research, Bangalore - 560089, India}
\author{Manas Kulkarni}
\affiliation{International centre for theoretical sciences, Tata Institute of Fundamental Research, Bangalore - 560089, India}

\begin{abstract}  
We investigate and map out the non-equilibrium phase diagram of a generalization of the well known  Aubry-Andr\'e-Harper (AAH) model. This generalized AAH (GAAH) model is known to have a single-particle mobility edge which also has an additional self-dual property akin to that of the critical point of AAH model. By calculating the population imbalance, we get hints of a rich phase diagram.  We also find a fascinating connection between single particle wavefunctions near the mobility edge of GAAH model and the wavefunctions of the critical AAH model. By placing this model far-from-equilibrium with the aid of two baths, we investigate the open system transport via system size scaling of  non-equilibrium steady state (NESS) current, calculated by fully exact non-equilibrium Green's function (NEGF) formalism. The critical point of the AAH model now generalizes to a `critical' line separating regions of ballistic and localized transport. Like the critical point of AAH model, current scales sub-diffusively with system size on the `critical' line ($I\sim N^{-2\pm0.1}$). However, remarkably, the scaling exponent on this line is distinctly different from that obtained for the critical AAH model (where  $I\sim N^{-1.4\pm0.05}$). All these results can be understood from the above-mentioned connection between states near mobility edge of GAAH model and those of critical AAH model. A very interesting high temperature non-equilibrium phase diagram of the GAAH model emerges from our calculations. 
\end{abstract}

\maketitle

\textit{Introduction:}   
Anderson localization is a phenomenon seen in a wide class of systems \cite{Mirlin_Evers,Physics_Today_1,Physics_Today_2}. It refers to spatial localization of energy eigenstates  in the presence of uncorrelated disorder and in absence of interactions. In one and two dimensions, even a small amount of disorder makes all energy eigenstates localized.   In three dimensions, beyond a critical strength of disorder, there occurs a mobility edge \cite{me0} separating localized and extended eigenstates. Understanding the physics of such a three dimensional system from a microscopic model is difficult. As a result, it is of interest to develop and study, theoretically and experimentally, lower dimensional models with a mobility edge.

One way to reproduce some of the physics of higher dimensional disordered systems in lower dimensions is to replace the uncorrelated disorder by a quasiperiodic potential. Such lower dimensional systems can also be connected to higher dimensional systems in presence of a magnetic field (like a quantum-Hall set-up) {\cite{Hofstader,Zilberberg2012,Zilberberg2013}. Such models are not only of interest in Physics, but are also studied in Mathematics \cite{math1,math2}.  A paradigmatic example of such a system in one-dimension is the Aubry-Andr{\'e}-Harper (AAH) model \cite{aa1,harper}. In the AAH model, with increasing strength of the quasiperiodic potential, there occurs a phase transition from all states being delocalized to all states being localized (and hence no mobility edge). The transition occurs through a critical point. Via a transformation (akin to a Fourier transform) a dual to the AAH model can be obtained, where localized and delocalized regimes are interchanged. At the critical point, the AAH model is self-dual under this transformation \cite{aa1}. The eigenstates at the critical point are neither totally delocalized nor localized, but are `critical' \cite{pandit83}. The transport properties at the critical point of AAH model has been studied recently in \cite{ap1}, and subsequently in \cite{vkv}. Sub-diffusive scaling of current with system size has been observed at the critical point. 

\begin{figure}
\includegraphics[height=5cm,width=\columnwidth]{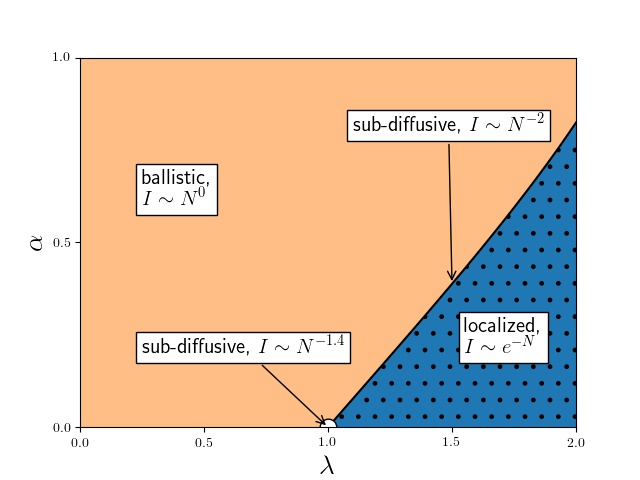} 
\caption{(color online)  Non-equilibrium  phase diagram at high temperatures for the GAAH model obtained from system size scaling of NESS current.  Here $I$ is NESS current and $N$ is the system size. } 
\label{fig:phase_diag}
\end{figure}

The AAH model and its generalizations have been of growing interest recently owing to their experimental realization in various set-ups  and the possibility of observing many-body localization and topological effects in such set-ups \cite{expt0,expt1,expt2,expt3,Zilberberg2013,Zilberberg2013_2,expt4,expt5,expt6}. Even though the AAH model has no mobility edge, various generalizations of it, as well as other quasiperiodic systems, have been  shown to have mobility edges in one dimension. Recently, physics of such systems have attracted a lot of attention \cite{AAH1, AAH2, AAH9,AAH10, AAH11,Nakayama2015,comment2015,garg,bi1,bi2,sg3,sg1,tridev} and has also been experimentally realized \cite{Bloch_mob_edge}. In this paper, we focus on one such model that has been recently proposed \cite{AAH1, AAH2, AAH9,AAH10, AAH11}. This model has a mobility edge, with the additional property that the mobility edge  is a self-dual point under a similar transformation as in the conventional AAH model. Henceforth, we call this model the generalized Aubry-Andr{\'e}-Harper (GAAH) model.

Although the phase diagram of AAH model is well-known, the phase diagram of GAAH model has not been studied. In general, non-equilibrium phase transitions have  received a lot of limelight recently, appear in various contexts \cite{np0,np1,np2,np3}. In this Letter, we map out the high temperature non-equilibrium phase diagram of the GAAH model (see Fig. \ref{fig:phase_diag}).  To this end, we first show that the evolution of particle density starting from different initial conditions, like only even sites occupied (population imbalance) and a step profile, gives hints of a rich phase diagram. We then describe a surprising correspondence between wavefunctions of GAAH model near mobility edge and those of the critical AAH model. As a consequence, a `critical line' in GAAH model separating regimes of ballistic and localized transport is found. For parameters on this `critical line', open system transport is sub-diffusive, but with an exponent different from that of critical AAH model.

\begin{figure}
\includegraphics[height=7cm,width=\columnwidth]{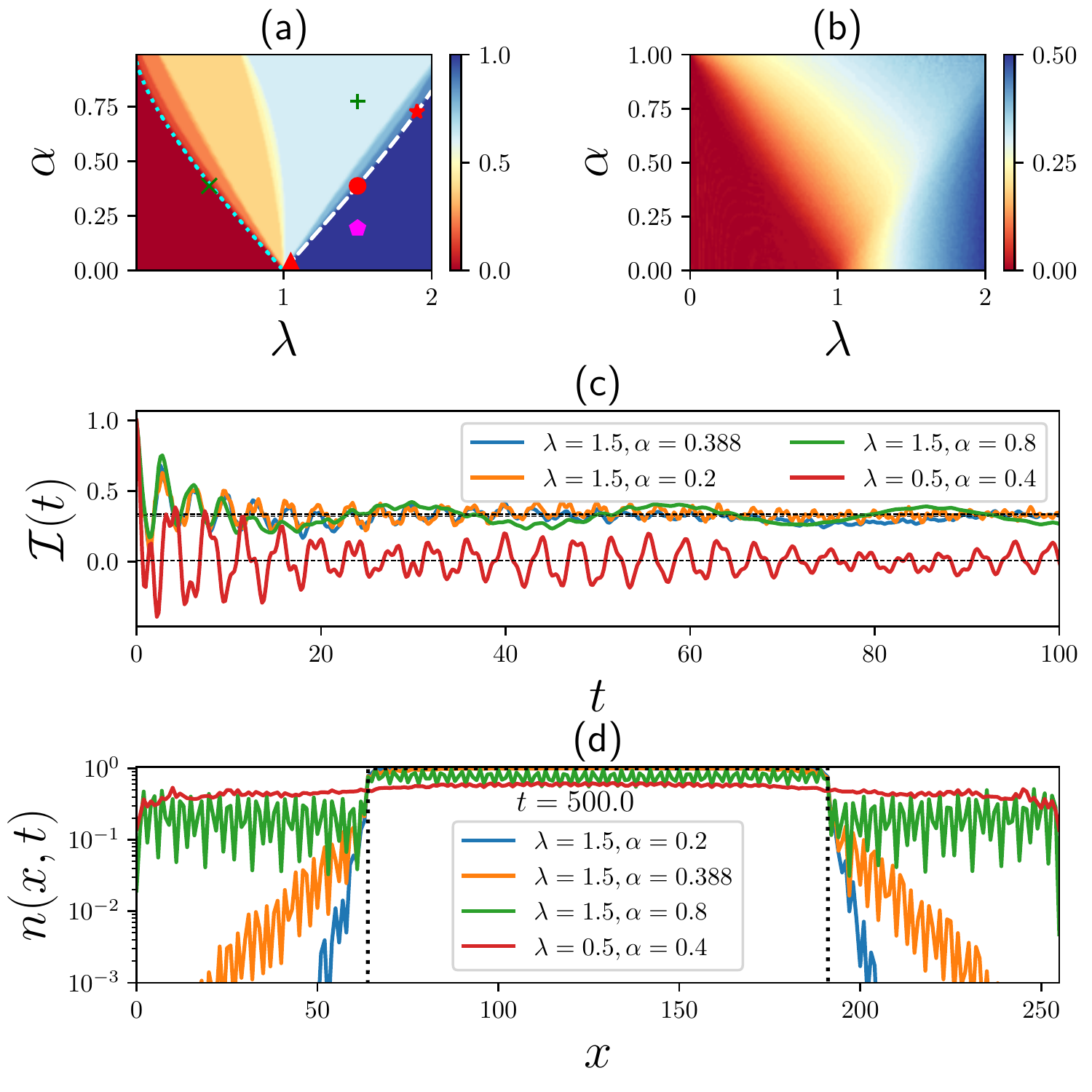} 
\caption{(color online)\textbf{(a)} Fraction of localized eigenstates (color coded) as a function of $\alpha$ and $\lambda$. A possible `phase diagram' with phase transitions and crossovers is evident.  The marked points are considered for further analysis. \textbf{(b)} The long time value of imbalance $\mathcal{I}$ (color coded) as a function of $\alpha$ and $\lambda$ maps out a `phase diagram' quite similar to that in (a).  \textbf{(c)} Behavior of $\mathcal{I}$ with time for some chosen points. (d) A long time snap-shot result of an initially localized profile (shown with the dotted line) for various values of $\lambda$ and $\alpha$.   Parameters : $N=1024$ for \textbf{(a)}, $N=256$ for \textbf{(b)}, \textbf{(c)} and \textbf{(d)}.}
\label{fig:frac_loc_imbalance}
\end{figure}

\begin{figure}
\includegraphics[height=7cm,width=\columnwidth]{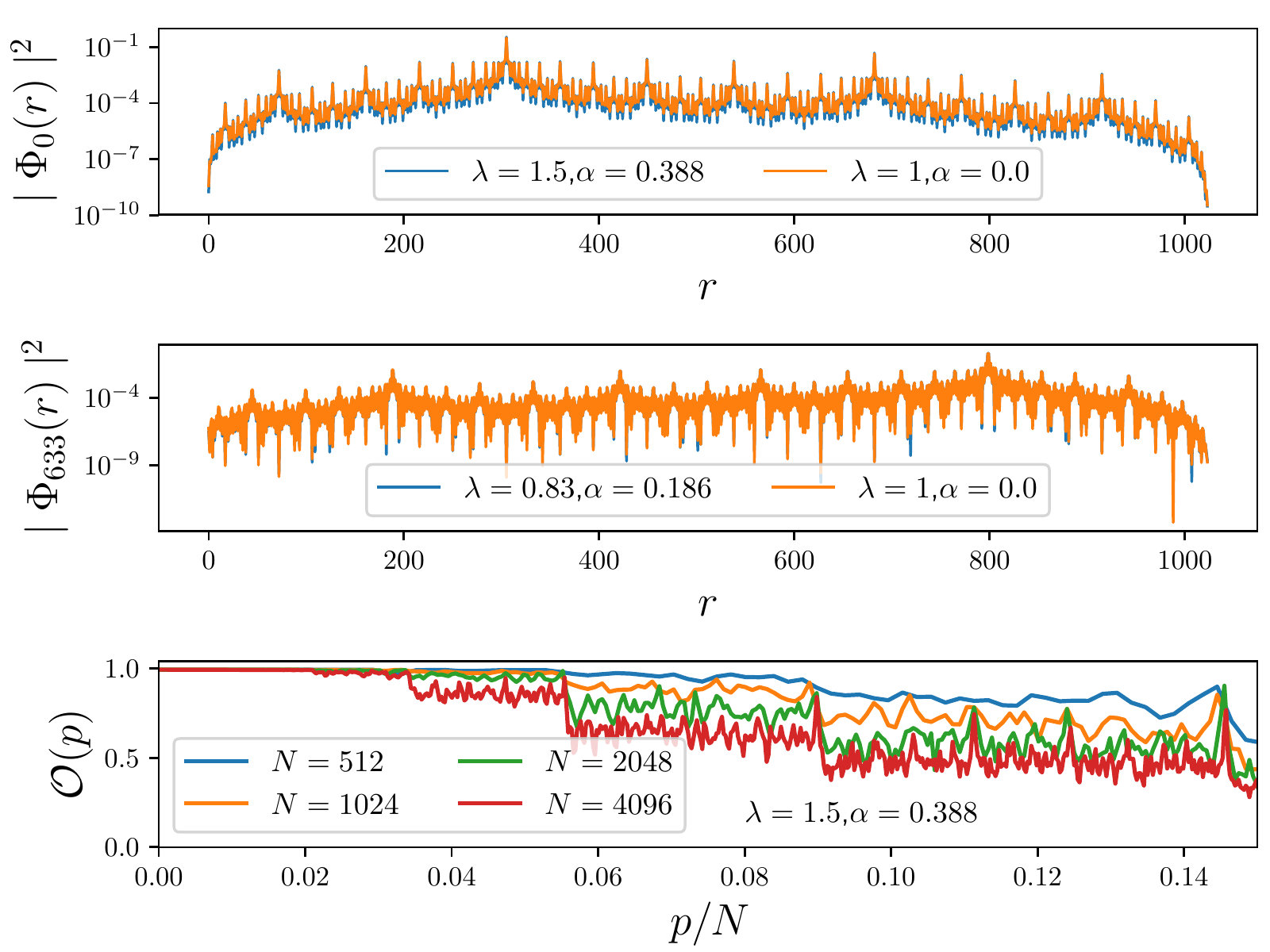} 
\caption{(color online) Top two panels: The figure demonstrates a remarkable matching of single particle wavefunctions $\Phi_{p}(r)$ of the GAAH model near the self-dual point with those of critical AAH model ($\alpha=0$, $\lambda=1$).  Bottom panel: Overlap `integral' of wavefunctions of GAAH and AAH models as function of eigenstate index for several system sizes. Here $p$ is the eigenstate index and $r$ is the site index. Parametes: Top two panels: $N=1024$, $\phi=0$. Bottom panel is averaged over $\phi$.}
\label{fig:wavefunc}
\end{figure}

\textit{Model:} 
The generalized AAH model (GAAH) is given by the Hamiltonian 
\begin{align}
&\mathcal{H}_S~=~\sum_{r=1}^{N-1}(\hat{a}_r^{\dagger} \hat{a}_{r +1}+h.c)+\sum_{r=1}^N   \frac{2\lambda \cos(2\pi b r+\phi)}{1-\alpha \cos(2\pi b r+\phi)}\hat{a}_r^{\dagger} \hat{a}_r \label{H_S},
\end{align}
and $\{\hat{a}_r\}$ are fermionic (bosonic) annihilation operators, 
$\lambda$ is the strength of the onsite potential, $\phi$ a phase factor, $b$ is a an irrational number and  the parameter 
 $\alpha \in (-1,1)$.  For our analysis, we will restrict ourselves to $\alpha,\lambda>0$. Other regions can be reconstructed via symmetries in the Hamiltonian. The symmetries can be concisely written as $\mathcal{H}_S(\lambda,\alpha,\phi) = \mathcal{H}_S(-\lambda,-\alpha,\phi+\pi)=-\tilde{\mathcal{H}}_S(-\lambda,\alpha,\phi)=-\tilde{\mathcal{H}}_S(\lambda,-\alpha,\phi+\pi)$, where $\tilde{\mathcal{H}}_S$ is the Hamiltonian obtained after the transformation $\hat{a}_r \rightarrow (-1)^r \hat{a}_r$.  
For $\alpha=0$, we get the AAH model, where it is known that $\lambda=1$ is a critical point exhibiting self-duality.  Similarly it is known that for any choice of $b$ and $\phi$, the GAAH model shows self-duality at an energy $E$ satisfying the condition  $\alpha E = 2\, \mbox{sgn}(\lambda)(1 -\mid \lambda \mid)$,  provided $E$ is a single-particle energy eigenvalue. All energy eigenstates  with energy less than $E$ are extended while those higher than $E$ are localized. If $E$ falls within the spectrum, then it is a mobility edge. In this paper, we investigate the phase diagram of this system in the $\alpha$-$\lambda$ plane.
We choose $b=(\sqrt{5}-1)/2$, which is the golden mean, and, unless otherwise mentioned, all our results are obtained after averaging over $\phi$. 

The first hint of a rich phase diagram of this model comes from direct calculation of the fraction of localized states in the system. It can be checked via calculation of inverse participation ratio (IPR) that, for $\lambda,\alpha>0$, the states with energy greater than the self-dual point are localized. Thus, the fraction of localized states is given by the fraction of single particle eigenstates with energy greater than the self-dual point. This is shown in Fig.~\ref{fig:frac_loc_imbalance}(a). We immediately see several regions  of different colors (indicating different fraction of localized states) with clear boundaries. Two of the most clear boundaries are shown by the  dotted and dashed lines. Below the dotted line, there is no mobility edge and all states are delocalized. This line corresponds to the case where the self-dual point exactly coincides with the highest energy eigenstate of the system. Below the dashed line, there is no mobility edge and all states are localized. This line corresponds to the case where the self-dual point exactly coincides with the lowest energy eigenstate of the system. As we will see below, this is actually a `critical' line in the non-equilibrium phase diagram. Henceforth, we will call this line the `critical' line of GAAH model. Note that since the minimum energy eigenvalue depends on the choice of irrational number, the `critical' line of GAAH model is not independent of choice of irrational number.

Next, we check if a physically measurable quantity can reproduce all the regions shown by the fraction of localized states. To this end, we calculate the population imbalance. This is defined by $\mathcal{I}(t)=\frac{\langle N_e(t) - N_o(t) \rangle_0 }{ \langle N_e(t) + N_o(t) \rangle_0}=\frac{2}{N}\sum_r (-1)^r\langle  \hat{n}_r(t) \rangle_0 $, where $N_e$ and $N_o$ are the number of particles at the even and odd sites respectively and $\langle ... \rangle_0$ denotes expectation value over the initial state. The initial state is chosen so that only the even sites are occupied. This quantity has been recently experimentally measured for the regular AAH model \cite{expt0,expt1,expt2}. $\mathcal{I}(t)$ tends to a steady value in the long time limit (Fig.~\ref{fig:frac_loc_imbalance}(c)). This value depends monotonically on the number of localized states.  Hence variation of long time value of $\mathcal{I}(t)$ with $\alpha$ and $\lambda$, should be similar to the fraction of localized states plot. This is exactly what we find in  Fig.~\ref{fig:frac_loc_imbalance}(b) which shows the long time value of $\mathcal{I}(t)$ (color coded) as a function of $\alpha$ and $\lambda$. However, the `critical' line is not well captured via $\mathcal{I}(t)$. This is because on this line there are still a large number of localized states. While imbalance captures the presence of localized states, it cannot directly capture the effect of having a mobility edge. This is nicely captured by evolution of an initially localized step particle density profile \cite{moore} (Fig. \ref{fig:frac_loc_imbalance}(d)). Delocalized states cause the initially localized profile to spread out with time, while localized states almost do not evolve the initial profile. In presence of a mobility edge, there is a coexistence of both kinds of behaviors due to presence of both localized and delocalized states. This has recently been seen in experiment \cite{Bloch_mob_edge}.

When there is a state at the mobility edge, it is a self-dual point. However this does not say anything about the nature of  energy eigenstates at or very close to the self-dual point. To check the nature of these states, we plot the
 wavefunctions of the GAAH model near the self-dual point. Let the eigenstates be ordered in ascending order of energy. Then, quite surprisingly, we find that, if the $p$th state of GAAH model is near the self-dual point, then its wavefunction almost exactly overlap with the $p$th state of the critical AAH model ($\alpha=0$, $\lambda=1$). This remarkable result immediately establishes that states near the self-dual point  have a `critical' nature. This phenomenon is observed not only on the `critical' line, but anywhere in the $\alpha$, $\lambda$ plane where there are eigenstates close to the self-dual point (Fig.~\ref{fig:wavefunc}). To quantify this, we calculate the overlap `integral' of the GAAH model and AAH model wavefunctions as a function of eigenstate index $p$, $\mathcal{O}(p)~=~\mid~\sum_{r=1}^N\Phi_p^{GAAH}(r)\Phi_p^{AAH}(r)~\mid$. This is shown in Fig.~\ref{fig:wavefunc} bottom panel for parameters on the `critical' line. Close to the minimum eigenvalue there is near complete overlap ($>99\%$). The plot also shows that, for a given system size, there is a finite fraction of such `critical' states, and the fraction of such states goes down with increase in system size. Note that although it has been previously shown that wavefunctions of GAAH model near self-dual point and those of critical AAH model have same IPR scaling exponents \cite{AAH2}, that does not say anything about the spatial overlap of wavefunctions. For example, two eigenstates of the critical AAH model have the same IPR scaling exponents but have zero overlap due to orthogonality. On the other hand, knowing that two states are overlapping automatically establishes them to have same scaling exponents. In this sense, the overlap we observe here, is a more direct and much stronger statement about the GAAH-AAH model wavefunction correspondence.

On the `critical' line of GAAH model, there are no truly delocalized states, but critical and localized states. Hence, contribution to transport properties should primarily come from these critical states. Since the wavefunctions of these states almost exactly overlap with critical AAH model, one can expect the transport properties on the `critical' line to be similar to that of the critical AAH model.  
We explore the transport behavior of GAAH model via scaling of current with system-size. To this end, we couple the fermionic (bosonic) system Hamiltonian $\mathcal{H}_S$ (Eq.~\ref{H_S}) bilinearly with two fermionic (bosonic) baths at two ends. The baths are modelled by non-interacting Hamiltonians with infinite degrees of freedom.   
The non-equilibrium steady state (NESS) current is given via the NEGF formula $
I = \int \frac{d\omega}{2\pi} \mathcal{T}(\omega)\left(n_1(\omega)-n_N(\omega)\right)$, where 
 $\mathcal{T}(\omega) =  \frac{J_1(\omega)J_N(\omega)}{\mid det(\mathbf{M}(\omega)) \mid^2} 
$, $n_1(\omega)=[e^{\beta_1(\omega-\mu_1)}+1]^{-1}$, $n_N(\omega)=[e^{\beta_N(\omega-\mu_N)}+1]^{-1}$ are the fermi-distribution functions with inverse temperature $\beta_1$ ($\beta_N$) and chemical potential $\mu_1$ ($\mu_N$) of the bath coupled to $1st$ ($N$th) site,  and $\mathbf{G}(\omega)=\mathbf{M}^{-1}(\omega)$ is the non-equilibrium Green's function (NEGF) of the system+bath. Writing the system Hamiltonian as $\mathcal{H}_S=\sum_{r,s}[\mathbf{H}_S]_{rs} {a}_r^ \dagger a_s$, $\mathbf{M}(\omega)$ is given by the $N\times N$ matrix $\mathbf{M}(\omega)~=~\left[ \omega\mathbf{I}- \mathbf{H}_S - \mathbf{\Sigma}^{(1)} (\omega)-\mathbf{\Sigma}^{(N)} (\omega)\right]$, where $\mathbf{\Sigma}^{(1)} (\omega)$, $\mathbf{\Sigma}^{(N)} (\omega)$ are   bath self energy matrices with the only non-zero elements given by  $\mathbf{\Sigma}^{(p)}_{pp}(\omega) = -\mathcal{P}\int \frac{d\omega^\prime J_p(\omega^\prime)}{2\pi(\omega^\prime-\omega)}-\frac{i}{2}J_p(\omega)$, $p=1,N$ ($\mathcal{P}$ denotes principal value). All integrals are over the entire spectrum of the baths. All information about the exact models of the baths is in the bath spectral functions $J_p(\omega)$. For the results presented here, we chose $J_1(\omega)=J_N(\omega)=J(\omega)=\frac{2\gamma^2}{t_B}\sqrt{1-\left(\frac{\omega}{2t_B}\right)^2}$. This corresponds to the baths being modelled by semi-infinite tight-binding chain with hopping parameter $t_B$ and bilinear system-bath coupling strength $\gamma$ \cite{ap1pra}. We have checked that our conclusions hold for other choices of bath spectral functions also (like Ohmic bath, i.e, $J(\omega)\propto \omega$,  etc.). Here we  present  results for the fermionic case under a chemical potential bias at a finite temperature (but no thermal bias), but the bosonic case and the thermal bias also lead to similar conclusions.


\begin{figure}[t]
\includegraphics[height=5cm,width=\columnwidth]{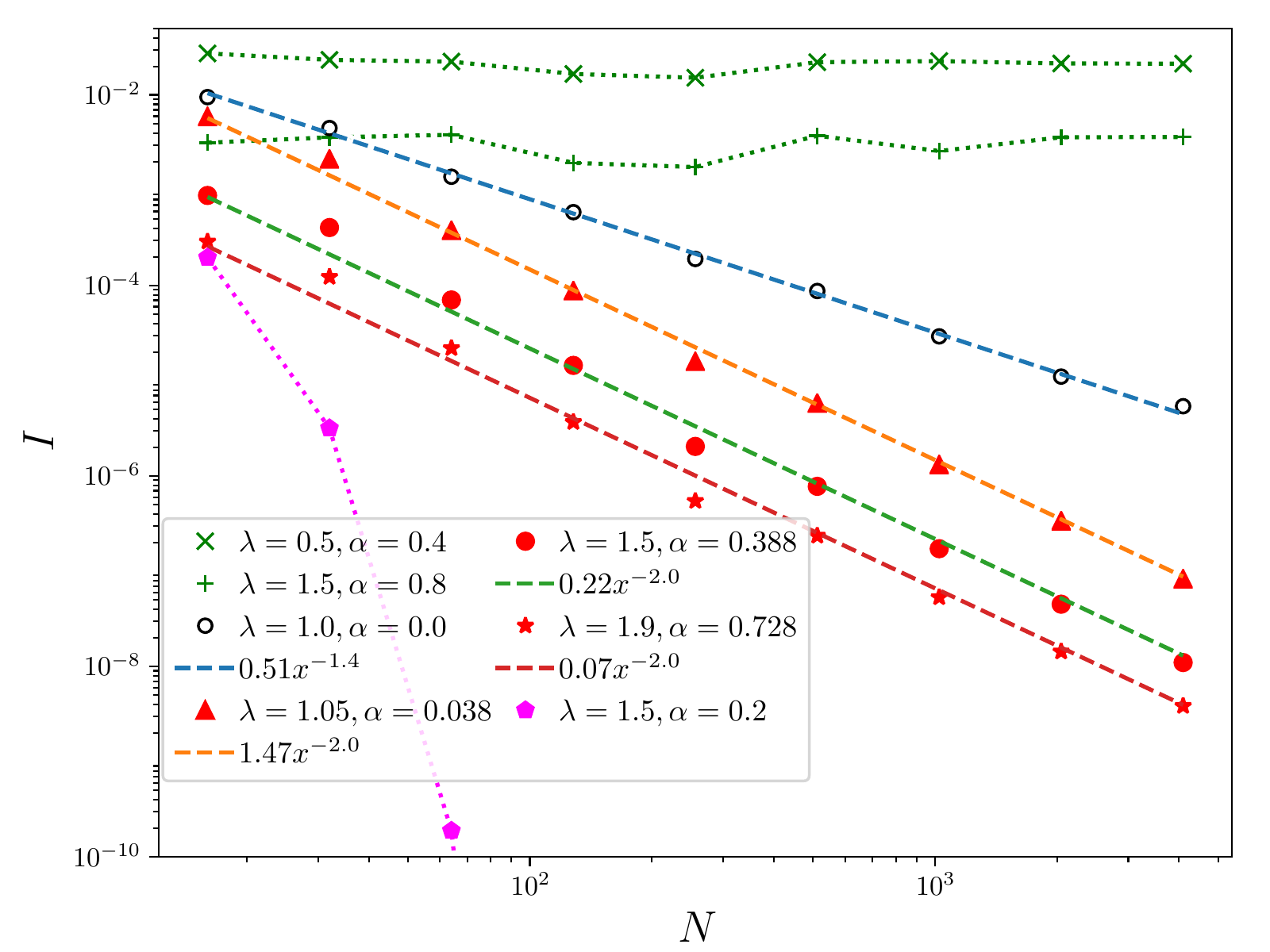} 
\caption{(color online)  The scaling of current with system size for various parameters of the GAAH model. On the `critical' line of the GAAH model, current scales as $I\sim N^{-2\pm0.1}$ showing sub-diffusive behaviour (red markers). In contrast, the critical AAH model ($\lambda=1.0,\alpha=0.0$) data shows $I\sim N^{-1.4 \pm 0.05}$ (white circles).  For parameters where some states are delocalized, transport is ballistic, $I\sim N^{0}$ (green markers). When all states are localized, current decays exponentially, $I\sim e^{-N}$ (magenta markers).  The points chosen in above plot correspond to the points in Fig.~\ref{fig:frac_loc_imbalance}(a) having the same symbol and color. The dashed lines are power law fits, whereas, the dotted lines are guide-to-eye.  Parameters : $\mu_1=50,\mu_2=-50,\beta=0.1,t_B=25,\gamma_1=\gamma_2=3$.}
\label{fig:current_scaling}
\end{figure}

The variation of current with system-size gives the nature of transport. The following two results are obvious: for parameters in the all-states-delocalized regime, the transport is ballistic and current is independent of system size ($I\sim N^0$) while for parameters in the all-states-localized regime, current decreases exponentially with system size ($I\sim e^{-N}$).  The interesting question is current scaling with system size in presence of mobility edge.
Fig.~\ref{fig:current_scaling} shows current $I$ as a function of system size for various values of $\alpha$ and $\lambda$. The most interesting finding is that, for parameters on the `critical' line, the $I\sim N^{-2\pm0.1}$, thereby showing transport is sub-diffusive. However, this is different from the critical AAH model  ($\alpha=0$, $\lambda=1$), where current scales as $I\sim N^{-1.4\pm0.05}$. For parameters where there are at least a few delocalized states, $I$ is independent of system size, which is a signature of ballistic transport (Fig.~\ref{fig:current_scaling}, bottom panel).  The above results are at relatively high temperature ($\beta=0.1$). At such temperatures, for any choice of chemical potentials $\mu_1$ and $\mu_2$, the same result will be seen.
Thus, the above results give us a clear non-equilibrium phase diagram of the GAAH model at high temperatures (Fig.~\ref{fig:phase_diag}).

To explain the difference in scaling exponent of $I$ vs $N$ between critical AAH model and `critical' line of GAAH model, we look at the system size scaling of the coarse-grained transmission near the minimum eigenvalue. For this, we first choose an energy range of interest and divide it into uniform cells of width $\delta E$. On this coarse-grained energy axis, the coarse grained transmission $\overline{\mathcal{T}}(\omega)$ is given by $\overline{\mathcal{T}}(E)=\left[\int_{E-\delta E/2}^{E+\delta E/2}\mathcal{T}(\omega) d\omega\right]/\delta E$.   We find that close to the minimum eigenvalue $\overline{\mathcal{T}}(E)~\sim~N^{-1.4\pm0.05}$, but the energy range upto which this scaling is seen decreases with increase in system size (Fig.~\ref{fig:T_and_f} left panel). This energy range corresponds to the energy range where GAAH and critical AAH model wavefunctions overlap. Since current is obtained by integrating over the transmission (with appropriate Fermi distributions), this directly gives a hint as to the reason for different exponent. The transmission, and hence $\overline{\mathcal{T}}(\omega)$, has peaks near system energy eigenvalues. Thus, we can approximate the integral over transmission as $\int\mathcal{T}(\omega)d\omega=\sum\overline{\mathcal{T}}(E)\delta E~\sim~N^{-1.4\pm0.05} (\frac{\Delta E}{\delta E})^{d_f}$, where $\Delta E$ is the energy range where states of GAAH model overlap with critical AAH model, and $d_f$ is the box-counting dimension of the spectrum over this energy range. $\Delta E \sim N^{-1.2\pm0.03}$ (Fig.~\ref{fig:T_and_f} inset). It can be checked that $d_f\sim 0.5$, which is the same as that of critical AAH model \cite{TangKohmoto1986}.  Thus $\int\mathcal{T}(\omega)d\omega$, and hence the current $I$, scales as $\sim N^{-2\pm0.1}$.

\begin{figure}[t]
\includegraphics[height=5cm,width=\columnwidth]{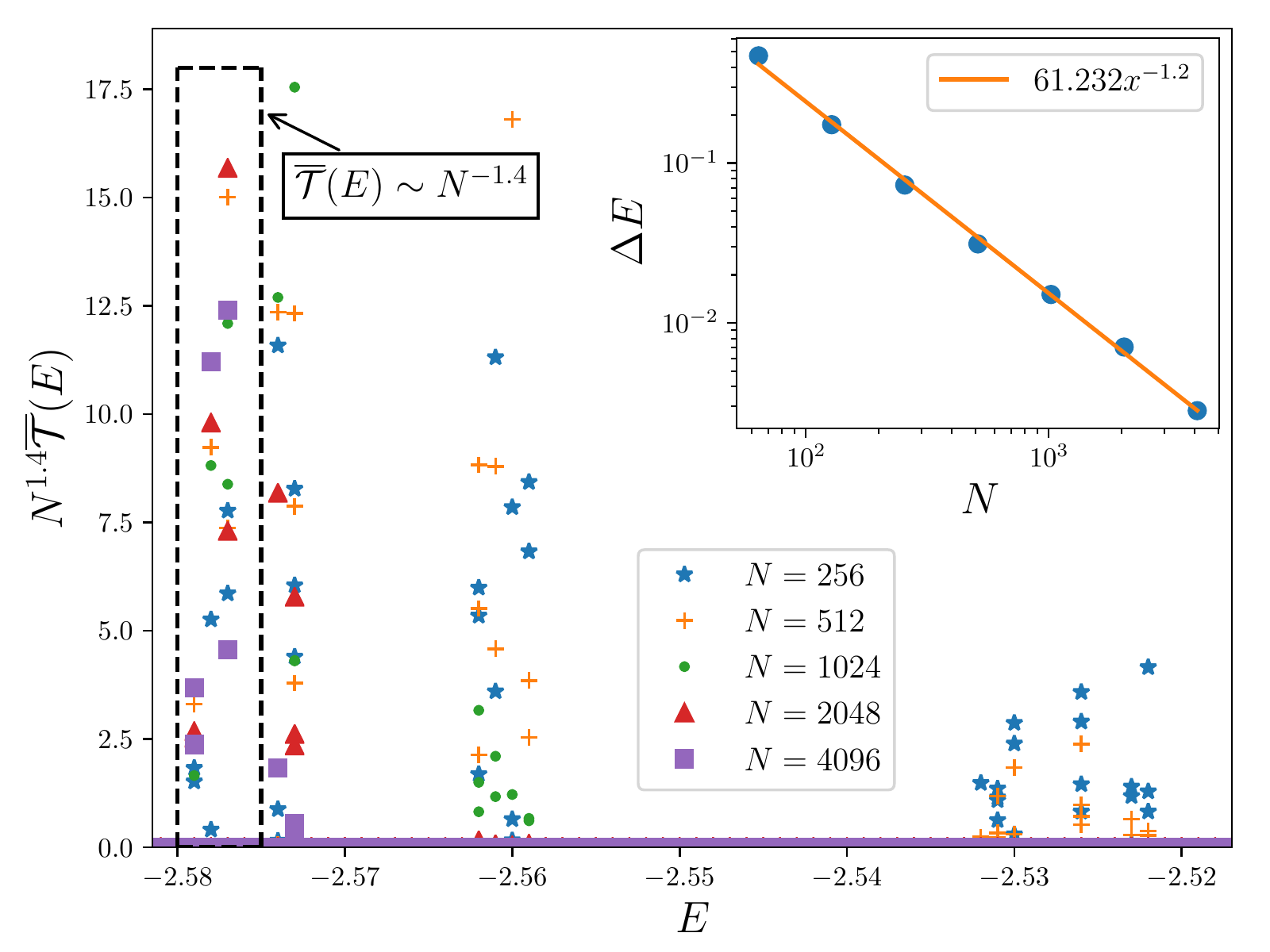} 
\caption{(color online) Coarse-grained transmission $\overline{\mathcal{T}}(E)$ near the self-dual point (the minimum energy eigenvalue) for a choice of parameters ($\lambda=1.5,\alpha=0.388$) on the `critical' line of GAAH model, scaled assuming similar scaling as critical AAH model. For larger system sizes, $\overline{\mathcal{T}}(E)\sim N^{-1.4\pm0.05}$ scaling holds over a smaller energy range. Inset: The plot of $\Delta E$ vs $N$. $\Delta E$ is the range of energy over which eigenstates of GAAH model that overlap more than $90\%$ with those of critical AAH model. $\Delta E\sim N^{-1.2 \pm 0.03}$.  Other parameters: $t_B=25,\gamma_1=\gamma_2=3,\delta E = 2\times 10^{-4}$. }
\label{fig:T_and_f}
\end{figure}

\textit{Conclusions:} Thus, in this letter, we have, for the first time, mapped out the high temperature non-equilibrium phase diagram of the GAAH model, which is a model with single particle mobility edge. In doing so, we have found a fascinating correspondence between the GAAH model and the conventional AAH model which does not have a mobility edge. It follows from this correspondence that the critical point of AAH model now generalizes to a `critical line' of GAAH model separating regions of ballistic and localized transport. However, the current scaling with system size on this `critical line' has a different exponent from that of the critical AAH model. We have also explained this from the GAAH-AAH model correspondence. The exact reason behind this correspondence is not clear, and requires further work which will be of great interest in mathematical front. Moreover, we would like to point out that, unlike the critical point of AAH model, the `critical' line of GAAH model depends on the choice of irrational number. The effect of other irrational numbers is also an interesting question. Also, it has been recently shown that the critical AAH model has remarkably different transport behaviors in closed and open system set-ups \cite{ap1}. Detailed investigation of closed system transport, as well as, low temperature transport of GAAH model is thus of great interest and will be taken up in a future work.

\textit{Acknowledgements: } We would like to thank Subroto Mukerjee and Sriram Ganeshan for useful discussions. M. K. gratefully acknowledges the hospitality of the Abdus Salam International Centre for Theoretical Physics (Trieste) during the conference on ``Many-Body-Localization: Advances in the Theory and Experimental Progress" where some interesting discussions took place. AD would like to thank support from the Indo-Israel joint research project No. 6- 8/2014(IC) and from the French Ministry of Education through the grant ANR (EDNHS). M. K. gratefully acknowledges the Ramanujan Fellowship SB/S2/RJN-114/2016 from the Science and Engineering Research Board (SERB), Department of Science and Technology, Government of India.

\bibliography{refGAA}
\end{document}